# LiDAR based determination of spring constant using smartphones

Sanjoy Kumar Pal[1], Soumen Sarkar[2], and Pradipta Panchadhyayee[3,4*]

[1]Anandapur H.S. School, Anandapur, PaschimMedinipur, West Bengal, India
[2]Karui P.C. High School, Hooghly, West Bengal, India
[3]Department of Physics (UG & PG), Prabhat Kumar College, Contai, PurbaMedinipur, India
[4]Institute of Astronomy, Space and Earth Science, Kolkata -700054, W. B., India
[*]E-mail: ppcontai@gmail.com

**Abstract**

A novel use of the LiDAR sensor of a smartphone in introductory physics experiments is discussed in this article. We have determined the spring constant for various combinations of springs using the LiDAR sensor of a smartphone through the phyphox application. An electrical heater coil is used as a spring, and the period of oscillation of a vertical spring-mass system is measured using a LiDAR sensor. The experimental values of spring constants agree with the theoretical values. A high school student can perform this simple experiment in a smart way at home.

## Introduction

With the improvement and inclusion of various types of sensors in smartphones, even relatively sophisticated science experiments can be performed at home. LiDAR sensor is one of those sensors available in high-end smartphones like iPad Pro 2020 and the iPhone 12 Pro. In 2020, Apple Inc. has released the phones mentioned above with the novel facility of having built-in LiDAR sensors. LiDAR, which stands for Light Detection and Ranging, is a remote sensing technology that uses laser light to measure distances and generate precise, three-dimensional information about the shape and characteristics of objects [1]. The scope of experiments using LiDAR is extensive and spans various fields. Here are some areas where LiDAR is commonly applied: Forestry, Geology, City planning, transportation, Archaeology, Surveying and Mapping, Meteorology and Climate Studies, Disaster management etc. The scope of experiments with LiDAR is dynamic and expanding as technology advances and researchers find new ways to leverage its capabilities in different fields [2]. With this aim, we have used the LiDAR sensor of iPhone12 pro max to design a new experiment on measuring the stiffness coefficient of springs.

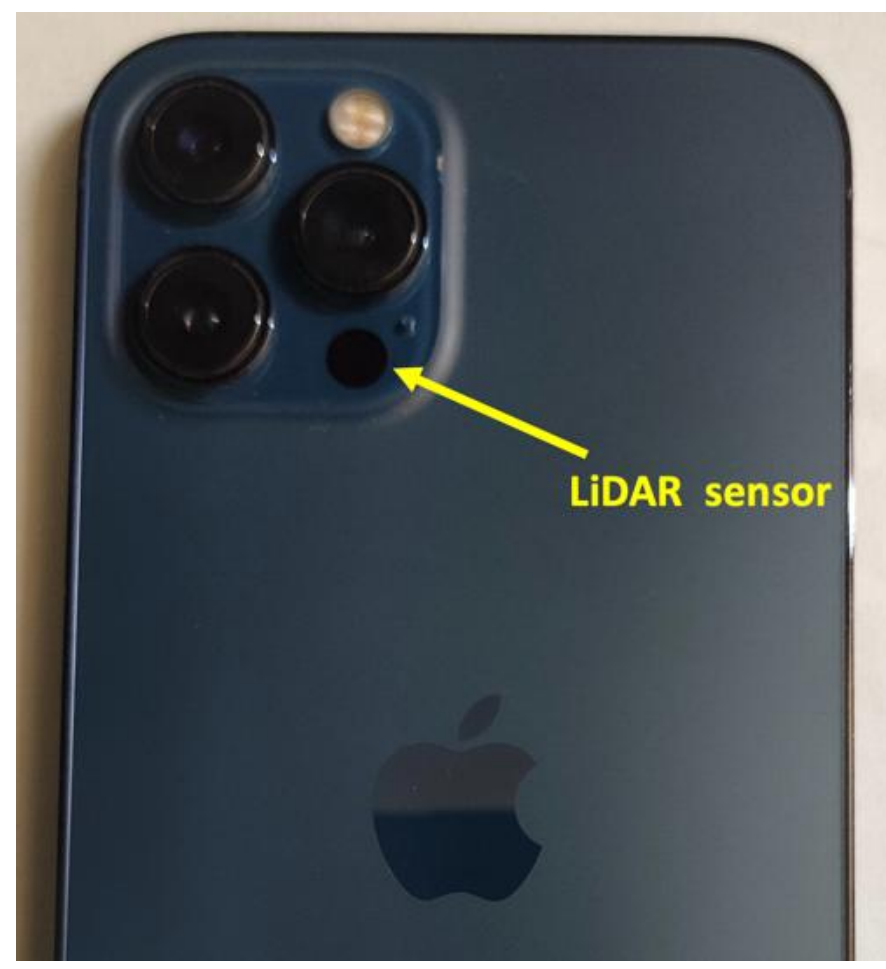

Fig. 1: LiDAR sensor in the smartphone, iPhone12 pro max.

Measurement of spring constant is a very common and easy experiment at school level. But the 'spring' concept is not restricted to this level. Rather, in condensed matter physics and quantum mechanics, the term 'spring' is often used in a metaphorical sense to describe certain types of restoring forces or interactions between particles. Some examples are: quantum harmonic oscillator, lattice vibrations (phonons), and magnetic springs (spin systems) etc. These metaphorical uses help the conceptualization of complex physical phenomena and make them more relatable. This spirit works behind the present work. Recently, experiments have been performed in this direction following various ways using smartphones [3-10]. In some works effective mass of the spring of a spring-mass oscillator is considered [3,8]. Series combination of springs has also been studied [9]. We have used a spring mass system as a vertical pendulum and noted the time period of oscillation using LiDAR through Phyphox application of iPhone12 Pro max. In this way the spring constant of the system is determined. We have verified the theoretical results concerned with different combinations of spring-mass systems - (i) series and (ii) parallel combinations, (iii) a special Y-type configuration with three springs.

**Theory**

The time period (*T*) of a spring-mass oscillator in the context of simple harmonic motion is given by the following equation:

$$T = 2\pi\sqrt{\frac{m+m_s}{k}}, \quad (1)$$

where *T* is the period of the oscillation, *k* is the spring constant, *m* is the mass of the suspended object, and $m_s$ is the effective mass of the spring.

Additionally, we mention that for the case $\frac{m}{m_{0s}} \gg 1$ ($m_{0s}$ is the mass of the spring), the effective mass $m_s$is given by: $m_s = \frac{m_{0s}}{3}$.

Equation (1) can be reframed as:

$$k = \frac{4\pi^2}{T^2} m_T, \quad (2)$$

where $m_T = m + m_s$.

This equation relates the square of the period and the stiffness constant (*k*) of the spring to the sum of the masses involved in the oscillation. Below, the equivalent spring constant for different spring systems are given.

(i) If two springs have spring constant $k_1$ and $k_2$, then the equivalent spring constant in series combination is $k_s$, where. $\frac{1}{k_s} = \frac{1}{k_1} + \frac{1}{k_2}$.

So, $k_s = \frac{k_1 k_2}{k_1+k_2}$. (3)

(ii) The equivalent spring constant in parallel combination is $k_p$ where $k_p = k_1 + k_2$. (4)

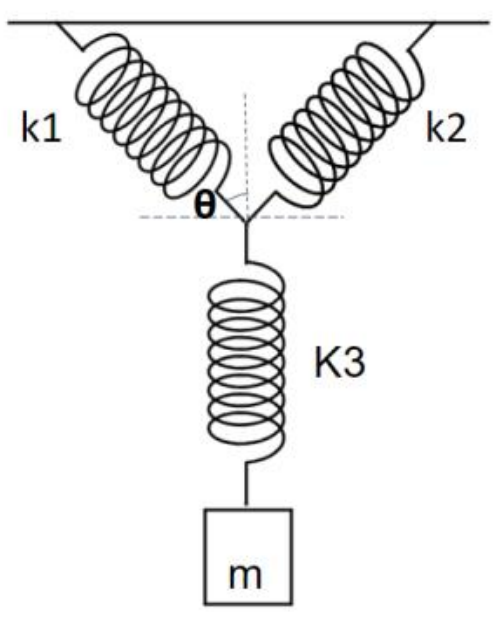

Fig. 2: Three springs are oriented in Y-shaped configuration.

(iii) In the Y-shaped configuration (Fig. 2) the spring-mass system consists of three springs of spring constants $k_1$, $k_2$, and $k_3$ where $k_1$ and $k_2$ are subtending an angle $\theta$ ( = 45°) with the vertical direction. The equivalent spring constant $k_{eq}$ is

$$k_{eq} = \frac{k_3(k_1+k_2)cos^2 45°}{(k_1+k_2)cos^2 45°+k_3}.$$

So, $$k_{eq} = \frac{k_3(k_1+k_2)}{(k_1+k_2)+2k_3}. \quad (5)$$

**Experimental setup and data**

Here, we have used a 1000W heater coil (Manufacturer: Heatco fire spiral) to determine the spring constant. The length of the coil is 41.0 cm and mass 20 g. The coil is suspended vertically from a fixed rod, and a hanger with variable mass is attached to the other end. A smartphone (iPhone 12 Pro Max) is positioned just below the system. To conduct the experiment, we utilize the LiDAR sensor of the iPhone. We launch the Depth sensor (LiDAR) within the Phyphox application and allow the spring-mass system to oscillate a couple of centimeters above the LiDAR sensor located on the backside of the smartphone. The LiDAR sensor of the iPhone emits invisible LASER radiation and receives the reflected rays to determine the depth or distance of an object. We have attached a 25 cm × 16 cm rectangular hard cardboard at the bottom of the cylindrical mass to enhance the sensitivity of the LiDAR sensor. To calculate the mass of the suspended object ($m$) we use the combined mass of the cylinder and the cardboard. The wave pattern observed on the Depth sensor (LiDAR) tab in the Phyphox application represents the oscillatory motion of the spring-mass system. The time versus distance graph obtained from the LiDAR sensor provides valuable information about the behavior of the system.

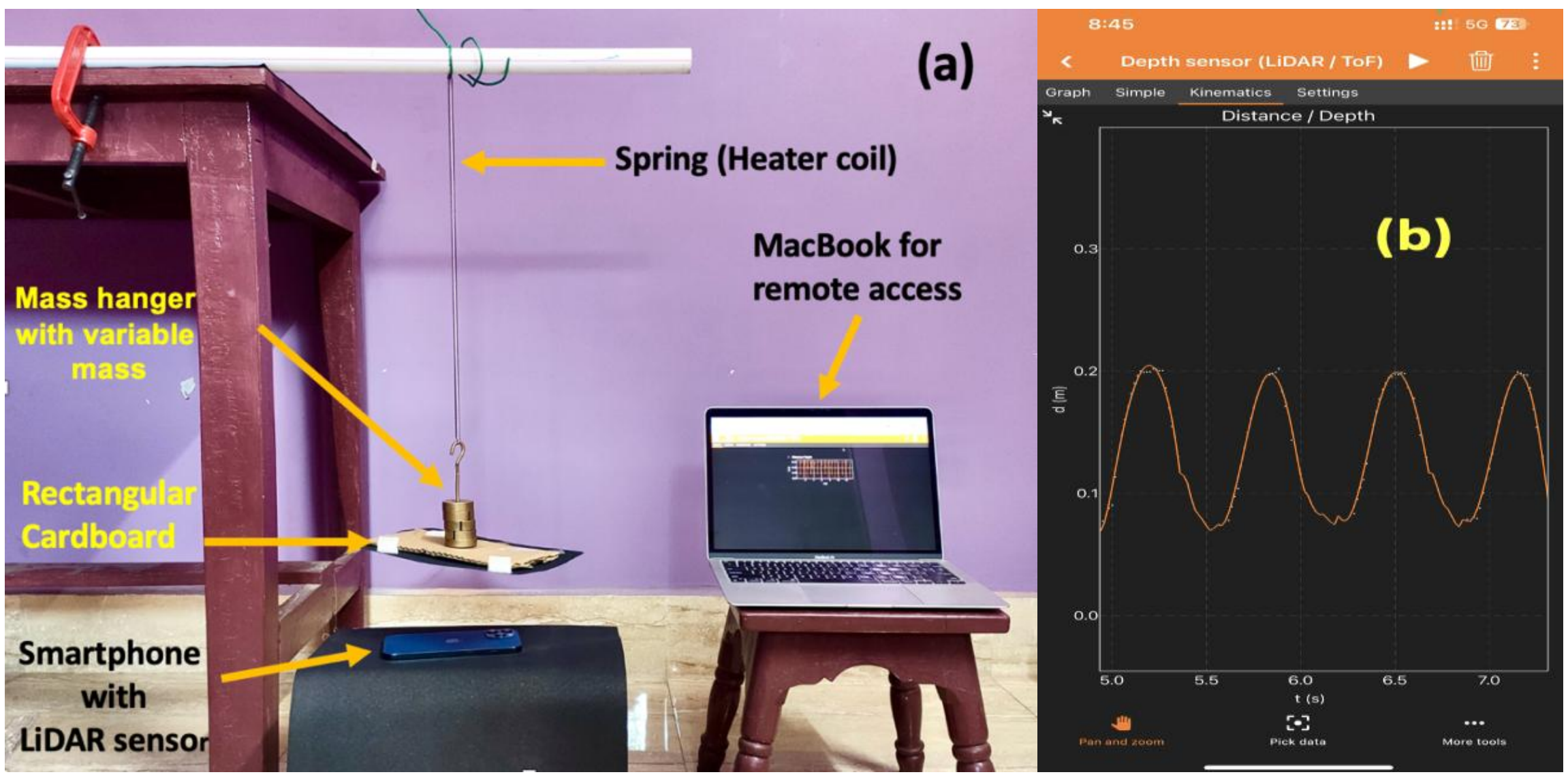

Fig. 3: (a) Experimental setup with a heater coil, a hanger with suspended mass, a rectangular cardboard , a smartphone and a laptop. (b) The time versus distance graph obtained from the action of LiDAR is shown in the Phyphox application window.

The graph displays the time versus distance pattern of the bottom of the mass-cardboard system. For better accuracy in determining the time period we allow 8 to 20 oscillations and note the total time taken from the data recorded by the LiDAR sensor. For the easier use of the application, we enable remote access to the Phyphox app and control the Depth Sensor (LiDAR) tab from a MacBook laptop connected to the same Wi-Fi as the smartphone being used. It is important to mention that the Depth Sensor on the smartphone will get activated once the smartphone is in position. This process ensures accurate data collection for the experiment. Make sure that the setup remains stable throughout the oscillations to obtain reliable results from the LiDAR sensor. The same process is repeated for determining the spring constant of various spring-mass combinations.

At the starting of data taking we determine the spring constant ($k$) of the entire coil. The corresponding mass ($m$) versus time period ($T$) data are presented in Table 1 and the $m - T^2$ graph is presented in Fig. 4.

Table 1: Table for Mass – Time period data for the entire coil

| Mass (kg) | Time-reading of the first* peak of oscillations (s) | Time-reading of the last* peak of oscillations (s) | Time difference between peaks (s) | No. of oscillations | Time period ($T$) (s) |
|---|---|---|---|---|---|
| 0.122 | 16.51 | 20.73 | 4.22 | 9 | 0.469 |
| 0.142 | 3.39 | 7.92 | 4.53 | 8 | 0.566 |
| 0.162 | 6.01 | 13.40 | 7.39 | 12 | 0.616 |
| 0.173 | 5.68 | 10.87 | 5.29 | 8 | 0.661 |
| 0.194 | 11.02 | 23.94 | 12.92 | 18 | 0.718 |
| 0.214 | 4.02 | 12.40 | 8.38 | 11 | 0.762 |
| 0.223 | 4.04 | 12.74 | 8.70 | 11 | 0.791 |
| 0.244 | 2.80 | 12.00 | 9.20 | 11 | 0.836 |
| 0.264 | 4.03 | 14.51 | 10.48 | 12 | 0.873 |
| 0.274 | 4.28 | 16.01 | 11.73 | 13 | 0.902 |
| 0.294 | 3.22 | 16.47 | 13.25 | 14 | 0.946 |
| 0.315 | 3.92 | 15.72 | 11.80 | 12 | 0.983 |
| 0.324 | 2.81 | 12.80 | 9.99 | 10 | 0.999 |

*Any peak during oscillation is considered as the first peak and the last peak is also selected after a finite number of oscillations.

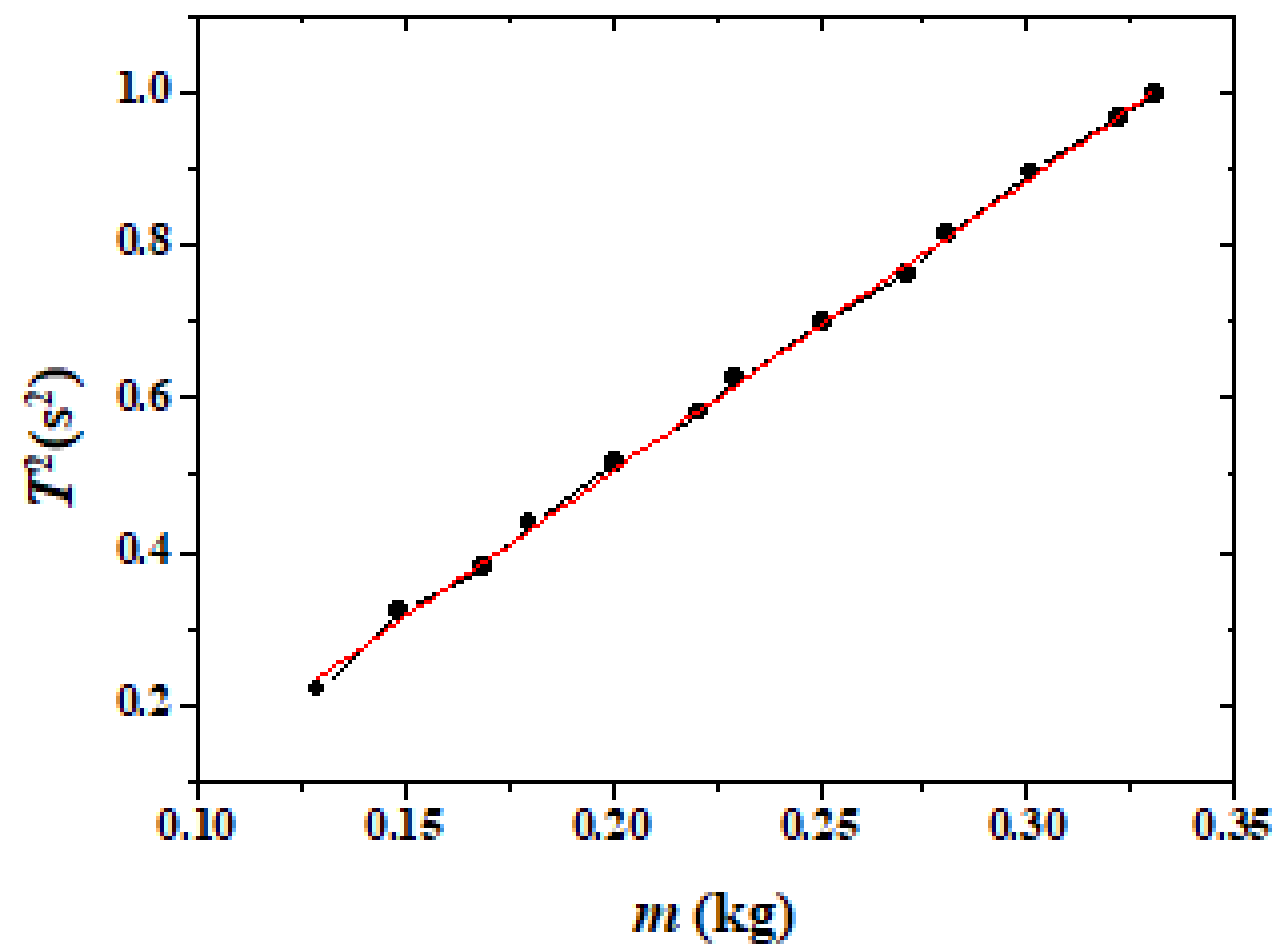

Fig. 4: $m - T^2$ graph for the entire coil.

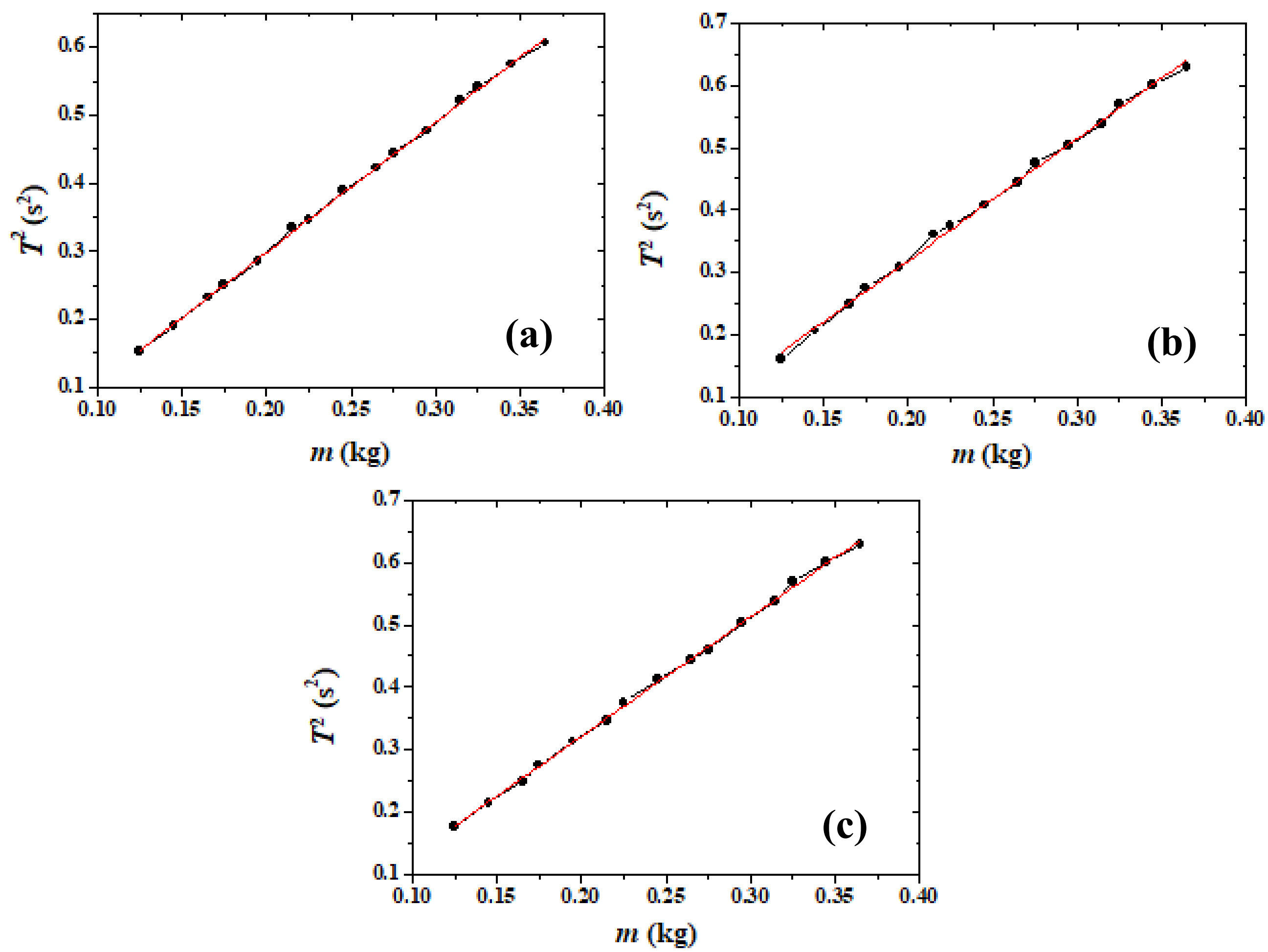

Fig. 5: $m - T^2$ graphs for (a) - A, (b) - B, and (c) - C springs.

For the verification of the established expressions (Eqs. (3-5)) for the equivalent spring constants we have taken two identical heater coils and cut each of them into two equal parts to make different combinations. Out of the four half-springs we have constructed the three springs designated as A, B, and C whose spring constants are designated as $k_1$, $k_2$, and $k_3$ , respectively. In the following, we have shown the $m - T^2$ graphs for A (spring constant: $k_1$), B (spring constant: $k_2$ ), and C (spring constant: $k_3$) springs.

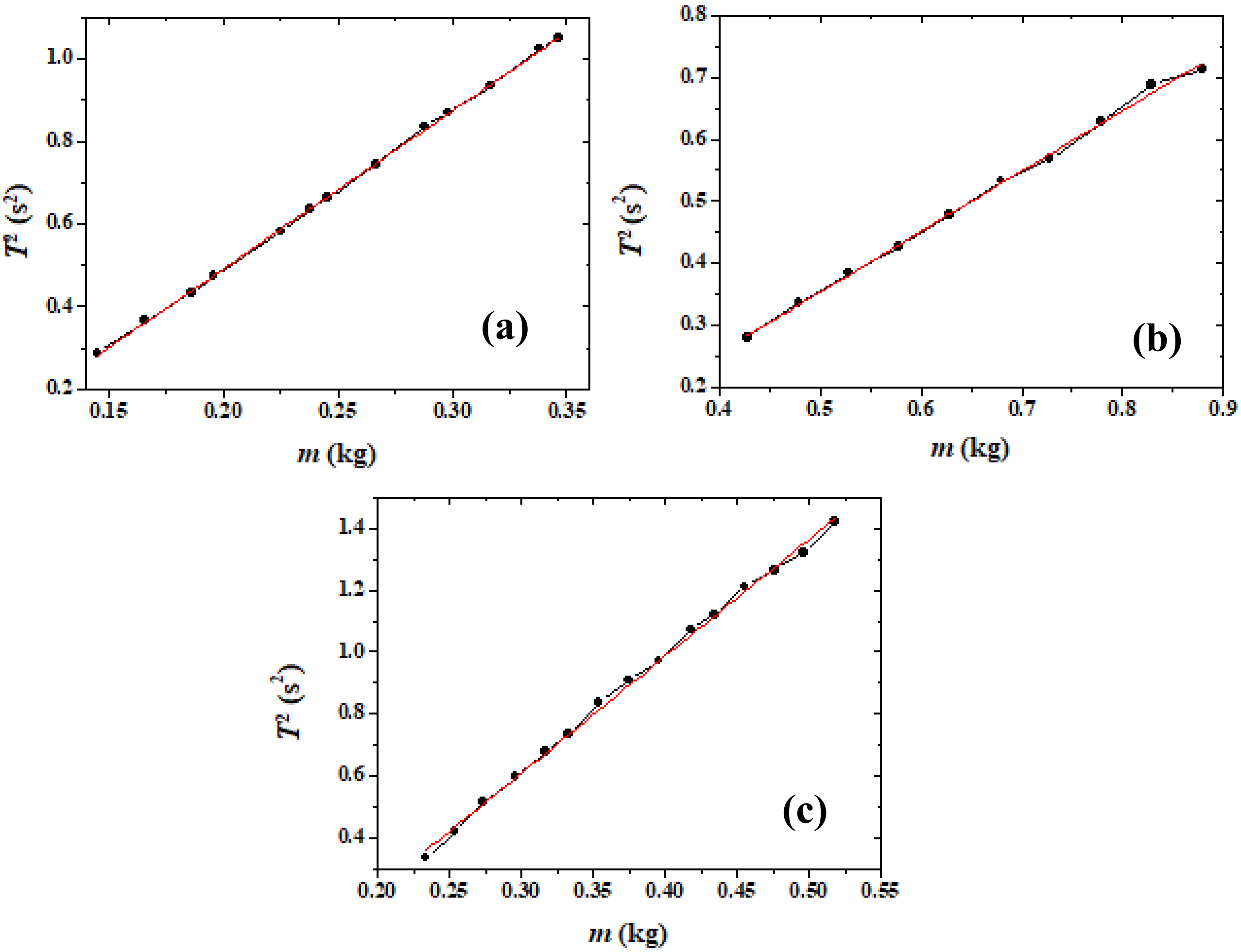

Fig. 6: $m - T^2$ graphs for (a) – series combination of A and B springs, (b) – parallel combination of A and B springs, (c) - Y-shaped configuration with A (left inclined), B (right inclined), and C (vertical) springs.

In the cases of the series and parallel combinations of spring-mass systems we have used A and B springs, while, for the Y-shaped configuration, we have used the spring C along with A and B springs. In this case, the angle of inclined springs with the vertical axis is determined by measuring the distance between supports and vertical distance of the junction point in equilibrium condition. Then measuring the time period, we have determined the spring constant for this combination. The final results of the values of spring constants are tabulated in the Table 2 with the calculated and measured values.

Table 2: Spring constants of the springs A, B, and C and their different combinations

| Spring | Slope ($s^2kg^{-1}$) | Value of spring constant ($Nm^{-1}$) (calculated) | Value of spring constant ($Nm^{-1}$) (measured) |
|---|---|---|---|
| Total spring (spring constant $K$) | 3.8002 (from Fig. 4) | 10.23 | 10.38 |
| Spring A (spring constant: $k_1$) | 1.9053 (from Fig. 5a) | - | 20.70 |
| Spring B (spring constant: $k_2$) | 1.9479 (from Fig. 5b) | - | 20.25 |
| Spring C(spring constant: $k_3$) | 1.9124 (from Fig. 5c) | - | 20.62 |
| Series combination (A & B) | 3.8191 (from Fig. 6a) | 10.23 | 10.33 |
| Parallel combination (A & B) | 0.9766 (from Fig. 6b) | 40.95 | 40.38 |
| Y-shaped configuration (A, B & C) | 3.7813 (from Fig. 6c) | 10.28 | 10.43 |

**Conclusion:**

We have measured the spring constants of the individual springs (nearly identical) and also determined the equivalent spring constants of the three different combinations (series, parallel, and Y-shaped configuration) of the springs. We have used LiDAR in measuring the spring constant. Table 2 shows that, for each combination, the measured values are very close to the calculated values. Note that the maximum error in determining the values of the spring constants is within 0.8% [*Detailed calculations are not shown here*]. This remarkable agreement of the theoretical and experimental values may encourage teachers to introduce this method of measurement of the spring constant of a spring-mass system in high schools.